\documentclass[aip,amssymb,jap,numerical,reprint]{revtex4-1}
\usepackage{graphicx}
\usepackage{dcolumn}

\begin{document}
\preprint{Proceedings for joint MMM/INTERMAG conf. 2013}

\title[Proceedings for joint MMM/INTERMAG conf. 2013]{Magnetic and transport properties of tetragonal- or cubic-Heusler-type Co-substituted Mn-Ga epitaxial thin films}

\author{T. Kubota}
\email[Author to whom correspondence should be addressed. Electronic mail: ]{takahide@wpi-aimr.tohoku.ac.jp}
\affiliation{WPI Advanced Institute for Materials Research, Tohoku University, Sendai 980-8577, Japan}
\author{S. Ouardi}
\affiliation{Department of Inorganic Chemistry, Max Planck Institute for Chemical Physics of Solids, Dresden 01187, Germany.}
\author{S. Mizukami}
\affiliation{WPI Advanced Institute for Materials Research, Tohoku University, Sendai 980-8577, Japan}
\author{G. H. Fecher}
\affiliation{Department of Inorganic Chemistry, Max Planck Institute for Chemical Physics of Solids, Dresden 01187, Germany.}
\author{C. Felser}
\affiliation{Department of Inorganic Chemistry, Max Planck Institute for Chemical Physics of Solids, Dresden 01187, Germany.}
\author{Y. Ando}
\affiliation{Graduate School of Engineering, Tohoku University, Sendai 980-8579, Japan.}
\author{T. Miyazaki}
\affiliation{WPI Advanced Institute for Materials Research, Tohoku University, Sendai 980-8577, Japan}

%\thanks{}

\date{\today}

\begin{abstract}
The composition dependence of the structural, magnetic, and transport properties of epitaxially grown Mn-Co-Ga films were investigated. The crystal structure was observed to change from tetragonal to cubic as the Co content was increased. In terms of the dependence of saturation magnetization on the Co content, relatively small value was obtained for the Mn$_{2.3}$Co$_{0.4}$Ga$_{1.3}$ film at a large {\it K}$_\textrm u$ value of 9.2 Merg/cm$^3$.
Electrical resistivity of  Mn-Co-Ga films was larger than that of pure Mn-Ga film. The maximum value of the resistivity was 490 $\mu\Omega$cm for Mn$_{2.2}$Co$_{0.6}$Ga$_{1.2}$ film. The high resistivity of Mn-Co-Ga might be due to the presence of localized electron states in the films due to chemical disordering caused by the Co substitution.
\end{abstract}


\maketitle

\section{INTRODUCTION}
Mn-Ga ordered alloys with tetragonal distortion, which are primarily used in spintronics applications, are known to possess high magnetic anisotropy; ({\it K}$_u$) \cite{Niida1996,Bal2007,Wu2009,Mizukami2012} this property of Mn-Ga alloys is essential ro the retention of stored data in spintronics-based memory devices with nanometer-scale elements. In addition, materials with high spin polarization\cite{Bal2007,Kurt2011,Kub2011} and small Gilbert damping constant ($\alpha$)\cite{Mizukami2011} are also particularly attractive for realizing spin-transfer-torque (STT)-type magnetoresistive random access memory (MRAM).\cite{Kish2008} The primary issue to be addressed in MRAM applications is to reduce the critical current ({\it I}$_c$) required to cause STT-induced magnetization switching. The current {\it I}$_c$ is proportional to the damping constant $\alpha$ and the saturation magnetization ({\it M}$_s$) of the free layer present in the magnetic tunnel junctions of the MRAM,\cite{Slon1996,Berg1996} and thus magnetic material with small {\it M}$_s$ value is important as well. {\it D}0$_{22}$-type Mn-Ga alloys have been known to exhibit relatively small {\it M}$_s$ value of about 250 emu/cm$^3$, and moreover Alijani {\it et al.} discovered that Co substitution of Mn in {\it D}0$_{22}$-type Mn$_3$Ga can reduce the {\it M}$_s$ value  further.\cite{Alijani2011} In a recent study, we obtained a high {\it K}$_u$ value of the order of 10$^7$ erg/cm$^3$ for an epitaxially grown {\it D}0$_{22}$ Mn-Co-Ga film.\cite{Ouardi2012} Our preliminarily work in the study demonstrated attractive possibilities of using such films for future memory applications; however, systematic investigations of epitaxially grown Mn-Co-Ga films over a wide composition range are still required. Therefore, we investigated structural, magnetic and transport properties of Mn-Co-Ga films in this study.

\section{EXPERIMENTAL}
All the films were prepared by using an ultra-high-vacuum magnetron sputtering system. The stacking structure of the samples was follows: MgO(100) substrate/Mn-Co-Ga (100 nm)/MgO(2 nm)/Al(2 nm). The Mn-Co-Ga layer was deposited via a co-sputtering technique from a Mn-Ga alloy target and an elemental Co target. The substrate was heated to 500$^\circ$C during deposition. In this study, primarily focused on the Co content dependences of structural, magnetic, and electric properties, and thus, the film compositions under investigations were Mn$_{2.6}$Ga$_{1.4}$, Mn$_{2.3}$Co$_{0.4}$Ga$_{1.3}$, Mn$_{2.3}$Co$_{0.5}$Ga$_{1.2}$, Mn$_{2.2}$Co$_{0.6}$Ga$_{1.2}$, Mn$_{2.1}$Co$_{0.8}$Ga$_{1.1}$, and Mn$_{1.8}$Co$_{1.2}$Ga$_{1.0}$; the Co content in the films was varied from 0 to 1.2.
The film compositions were determined by using inductively coupled plasma (ICP) mass spectroscopy for  Mn$_{2.3}$Co$_{0.4}$Ga$_{1.3}$, Mn$_{2.3}$Co$_{0.5}$Ga$_{1.2}$, and Mn$_{1.8}$Co$_{1.2}$Ga$_{1.0}$. The compositions of the other films were estimated via the ratio of the sputtering power used for each of the targets.
The structural and magnetic properties of the films were investigated by using an x-ray diffractometer (XRD) and a vibrating sample magnetometer (VSM), respectively. Electrical resistivity was measured using the van der Pauw technique.\cite{Pauw1958} All the experiments were performed  at room temperature.

 \begin{figure}
 \includegraphics[scale=0.85]{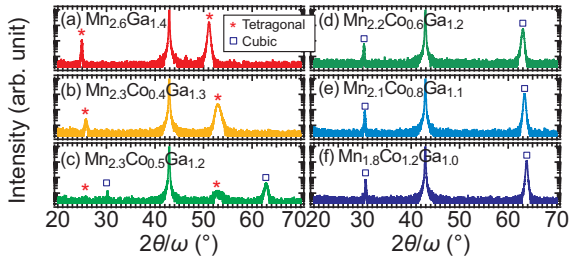}
 \caption{\label{XRD} (color online) Out-of-plane x-ray diffraction spectra of Mn-Co-Ga films with various composition ratios. Peaks marked with $\star$ and $\square$ represent the diffractions from the tetragonal phase and cubic phase, respectively. The large peak appearing at 2$\theta / \omega \sim$ 42$^\circ$ is originated from the (002) plane of the MgO substrate.}
 \end{figure}

\section{RESULTS and DISCUSSION}
\begin{figure}
 \includegraphics[scale=0.85]{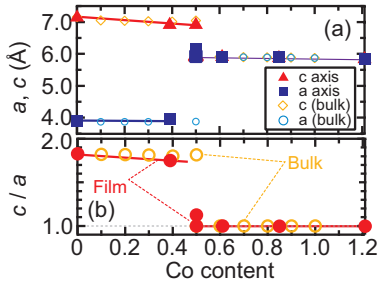}
 \caption{\label{lattice} (color online) (a) lattice constants ({\it a}- and {\it c}-axis) and (b) {\it c}/{\it a} ratio as a function of Co content, {\it x}. Lines are just guide to the eyes for data of the film samples.}
 \end{figure}

FIG. \ref{XRD} shows the out-of-plane XRD patterns of Mn-Co-Ga films. Only the peaks that originate from Mn-Co-Ga (002), (004), and MgO (002) planes appear in the XRD spectra. Peaks marked with $\star$ and $\square$ represent diffractions from the tetragonal ({\it D}0$_{22}$)\cite{Bal2007} and cubic ({\it X}$_a$, the so-called inverse Heusler)\cite{Helmholdt1987,Liu2008} structures, respectively. The epitaxial growth of the films and the superlattice diffractions of (011) (for {\it D}0$_{22}$) and (111) (for {\it X}$_a$) were confirmed via a $\phi$-scans for all the films (not shown here). The lattice constants ($\square$: a-axis, $\triangle$: c-axis) and the {\it c}/{\it a} ratio are summarized as a function of Co content of the Mn-Co-Ga films in FIGs. \ref{lattice} (a) and (b), respectively. The corresponding bulk values for Mn$_{3-{\it x}}$GaCo$_{{\it x}}$, as reported in ref. \onlinecite{Alijani2011} are also indicated by the open symbols. The Mn$_{2.6}$Ga$_{1.4}$ and Mn$_{2.3}$Co$_{0.4}$Ga$_{1.3}$ films exhibited a tetragonal structure while the Mn$_{2.3}$Co$_{0.5}$Ga$_{1.2}$ film exhibited both tetragonal and cubic phases; and the structure changed to cubic for films with larger Co content. The structural transition point of the film samples depending on the Co content is consistent with that for bulk ones even though the present Mn-Ga composition used in our study (Mn$_{2.6}$Ga$_{1.4}$) was different from the bulk one (Mn$_{3.0}$Ga$_{1.0}$), which implies that band dispersion is not very sensitive to the {\it off-stoichiometry} of the Mn-Co-Ga alloys; further tetragonal distortion occurs due to electronic instabilities corresponding to a band-type Jahn-Teller effect.\cite{Chad2012}

\begin{figure}
 \includegraphics[scale=0.85]{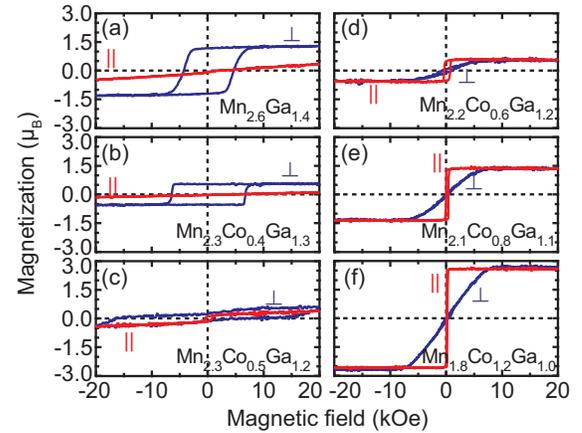}
 \caption{\label{MH} (color online) Hysteresis loops of Mn-Co-Ga films for various composition ratios. The magnetic field was applied perpendicular to the film plane direction for the curves indicated by $\perp$, and it was applied along the in-plane direction for those indicated by $\parallel$.}
 \end{figure}

FIG. \ref{MH} shows the hysteresis loops of the Mn-Co-Ga films. A magnetic field was applied perpendicular to the film plane direction for the curves indicated by $\perp$, and it was applied along the in-plane direction for those indicated by $\parallel$.
The hysteresis loops of the tetragonal samples show hard magnetic behavior with perpendicular anisotropy, while those of the cubic ones show soft magnetic behavior with in-plane anisotropy. The hysteresis loops of the Mn$_{2.3}$Co$_{0.5}$Ga$_{1.2}$ which contains both tetragonal and cubic structures exhibited undefined loop shapes of small values of magnetization.

\begin{figure}
 \includegraphics[scale=0.85]{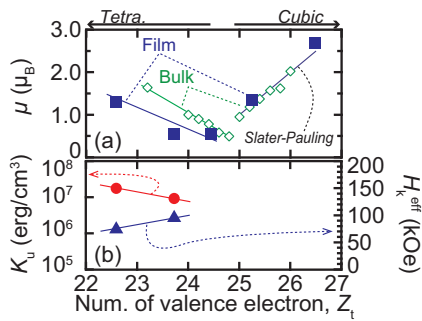}
 \caption{\label{m} (color online) (a) Magnetic moment ($\mu$) and (b) uniaxial magnetic anisotropy energy ({\it K}$_\textrm u$) and effective anisotropic field ({\it H}$_\textrm k^\textrm{eff}$) as a function of Co content in Mn-Co-Ga films. The lines serve as a visual guide to indicate the data curve for tetragonal samples. The lines for cubic samples indicates the Slater-Pauling rule.\cite{Kubler1984}}
 \end{figure}

The dependences of the magnetic moment ($\mu$), magnetic anisotropy energy ({\it K}$_\textrm u$), and effective anisotropy field ({\it H}$_\textrm k^\textrm{eff}$) as a function of the number of valence electron ($Z_t$) of Mn-Co-Ga films are shown in FIG. \ref{m}. The $\mu$ value of corresponding to bulk Mn$_{3-x}$GaCo$_x$ (as obtained from ref. \onlinecite{Alijani2011}) are also plotted in the figure.\cite{Note1} In addition, expected Slater-Pauling behavior\cite{Kubler1984} of Half-metallic Heusler compounds is also indicated for the cubic compositions. The values of {\it K}$_\textrm u$ were determined by using the relation {\it K}$_\textrm u$ = {\it M}$_\textrm s${\it H}$_\textrm k^\textrm{eff}$/2 + 2$\pi${\it M}$_\textrm s^2$ in the same manner as that described in our previous work.\cite{Mizukami2012} The dependence of $\mu$ for the film samples is similar to that of the reported bulk dependence. 
The $\mu$ exhibited minimum value around the boundary between the tetragonal and cubic structures. On the other hand, the value of {\it K}$_\textrm u$ and {\it H}$_\textrm k^\textrm{eff}$ did not widely differ for the samples with tetragonal structures. It is noteworthy that the magnetic moment ($\mu$) of the Mn$_{2.3}$Co$_{0.4}$Ga$_{1.3}$ film was as small as 0.55$\mu_\textrm B$ ($\sim$190 emu/cm$^3$) at a large {\it K}$_\textrm u$ value of 9.2 Merg/cm$^3$. In this case,  $\mu$ was reduced to less than half of that for the film without Co content, while a large {\it K}$_\textrm u$ close to 10 Merg/cm$^3$ was still maintained.

 \begin{figure}
 \includegraphics[scale=0.85]{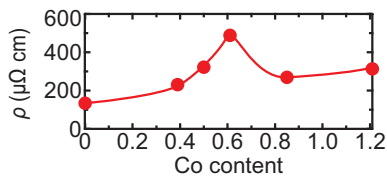}
 \caption{\label{r} (color online) Resistivity ($\rho$) as a function of Co content of Mn-Co-Ga films. The line serves as a visual guide.}
 \end{figure}

Subsequently, the electrical resistivity ($\rho$) of the Mn-Co-Ga films was investigated; the Co-content dependence of resistivity is shown in FIG. \ref{r}. The $\rho$ value of Mn$_{2.6}$Ga$_{1.4}$ was of the same order as that of Mn-Ga films with different composition ratio.\cite{Wu2010} With increasing Co content, $\rho$ correspondingly increased, and a maximum value of 490 $\mu\Omega$cm was obtained for the Mn$_{2.2}$Co$_{0.6}$Ga$_{1.2}$ film. As the Co content increased beyond 0.6, the $\rho$ slightly decreased; however it was still larger than that of 
pure Mn$_{2.6}$Ga$_{1.4}$. There are two possible explanations for the Co content dependence of $\rho$ for the Mn-Co-Ga films. The dependence could be extrinsic; films for Co content of around 0.5, the smoothness and continuity of the film become poor; for e.g., the roughness and peak-to-valley values of the Mn$_{2.2}$Co$_{0.6}$Ga$_{1.2}$ film were about 10 nm and 100 nm, respectively. These values were about 10 times larger than the corresponding value for the Mn$_{2.6}$Ga$_{1.4}$ film. Thus, increased electron scattering due to the presence of discontinuities in the film might be one reason for the observed $\rho$ behavior.
The other possibility is that the $\rho$ behavior is intrinsic to the chemical nature of the film. Chadov {\it et al.} have recently hypothesized that the chemical disordering of Mn$_3$Ga alloy by Mn-Co substitution can cause localization of the minority-spin channel around the Fermi level.\cite{Chad2012} They speculated that the electrical conductivity reduces because of the reduced mobility of electrons. According to their calculation, the localization is not sufficiently strong for the Mn$_3$Ga and Mn$_2$CoGa cases, while the electrons are strongly localized in both tetragonal and cubic Mn$_{2.5}$Co$_{0.5}$Ga compounds. In our study, the films with higher resistivity exhibit chemically disordered regions, and thus, such a localization can also be considered as the cause for the observed Co content dependence of $\rho$.

\section{SUMMARY}
Epitaxially grown Mn-Co-Ga tetragonal or cubic Heusler compound thin films were successfully fabricated, and their structural, magnetic, and electrical transport properties were investigated.
The dependences of lattice parameters and saturation magnetization on Co content in the films were similar to those reported for bulk Mn$_{3-x}$Co$_x$Ga compounds. A minimum saturation magnetization of 0.55$\mu_\textrm B$ ($\sim$190 emu/cm$^3$) was obtained for the Mn$_{2.3}$Co$_{0.4}$Ga$_{1.3}$ film at a large {\it K}$_\textrm u$ value of 9.2 Merg/cm$^3$; this magnetization value is acceptable in the light of future STT-MRAM applications. 
The resistivity of the Mn-Co-Ga films was larger than that for pure Mn-Ga. The observed higher resistivity subsequent to Co substitution might originate due to reduced electron mobility because of the presence of localized electron states around the Fermi level (intrinsic factor) as well as due to increase in surface discontinuities of the film (extrinsic factor).

\begin{acknowledgments}
This work was partly supported by the  ASPIMATT program (JST), Grant for Industrial Technology Research from NEDO, Grant-in-Aid for Scientific Research (JSPS), World Premier International Research Center Initiative (MEXT), and the Casio foundation.
\end{acknowledgments}


\end{document}